\documentclass[11pt]{article}

\usepackage{indentfirst}
\usepackage{amsmath}
\usepackage{amssymb}
\usepackage{amsfonts}
\usepackage[dvipsnames]{xcolor}
\usepackage{multirow}
\usepackage{makecell}
\usepackage{float}

\usepackage{geometry}
\usepackage[backend=bibtex, style=numeric-comp, sorting=none, maxnames=99]{biblatex}
\usepackage{hyperref}
\hypersetup
{
    colorlinks = true,
    linkcolor = red,
    citecolor = blue,
    urlcolor = PineGreen
}

\newcommand{\PR}{\scriptscriptstyle\mathrm{PR}}
\newcommand{\K}{\scriptscriptstyle\mathrm{K}}

\begin{document}

\begin{titlepage} \setcounter{page}{0}
\begin{center}
    \vspace*{1.0cm}
    {\Large\bf Matter Induced Bounds: Photon Ring, Lyapunov Exponent, Induced Temperature}
    \\ \vspace{1.0cm}
    {$\mbox{Hocheol Lee}$}\footnote{\it email: insaying@dongguk.edu}, \quad
    {$\mbox{Bogeun Gwak}$}\footnote{\it email: rasenis@dgu.ac.kr}
    \\ \vspace{0.2cm}
    {\small \it Department of Physics, Dongguk University, Seoul 04620, Republic of Korea}
    \\ \vspace{2.0cm}
\end{center}

\begin{center}
\begin{abstract}
    We investigate analytic bounds on the equatorial photon ring radius and the Lyapunov exponent in rotating black holes containing anisotropic matter. For a constant equatorial equation-of-state parameter, the finite asymptotic mass and nonnegative energy density imply an upper bound on the photon ring radius given by the Kerr photon ring radius with the same asymptotic mass and rotation parameters for co-rotating and counter-rotating motions. The lower bounds are written in terms of the quantities evaluated at the outer horizon. These bounds on the Lyapunov exponent constrain the absolute magnitude of the imaginary part of the quasinormal frequency of a massless scalar field within the eikonal limit. The worldsheet surface gravity and induced temperature were derived for a probe string in the Penrose limit along the photon ring. The induced temperature and worldsheet surface gravity give an upper bound on the photon ring Lyapunov exponent, $2 \pi T_\mathrm{ind} = \kappa_\mathrm{ind} \geq \lambda$.
\end{abstract}
\end{center}
\end{titlepage}

\newpage
\section{Introduction} \label{sec:introduction}
    Black holes lie at the intersection of the gravitational theory and astrophysical phenomena. Observations of black holes support key predictions of general relativity. The Event Horizon Telescope images of M87$^\ast$ and Sagittarius A$^\ast$ agree with predicted black hole shadows~\cite{EventHorizonTelescope:2019dse, EventHorizonTelescope:2022wkp}, while gravitational wave signals from binary black hole mergers match predicted coalescence waveforms~\cite{LIGOScientific:2016aoc}. The observed Schwarzschild precession of star S2 near Sagittarius A$^\ast$ gives an independent test for relativistic orbital motion~\cite{GRAVITY:2020gka}. Theoretically, the concept of black hole entropy~\cite{Bekenstein:1973ur} and the prediction that black holes emit thermal radiation because of quantum effects~\cite{Hawking:1975vcx} extend black holes outside classical general relativity to incorporate quantum and thermodynamic properties. Black holes link the gravitational theory with astrophysical phenomena, and including matter introduces additional questions regarding the geometric and physical features of black holes.

    Null geodesics model photon trajectories and are applied to interpreting black hole images. In black hole imaging, a photon ring comprises light rays reaching the observer after traversing the vicinity of unstable photon orbits. Near an unstable photon orbit, the growth of the small radial displacement is described by the Lyapunov exponent. The associated Lyapunov exponent per half orbit sets the exponential decrease of the successive photon subring radial widths on an observer screen~\cite{Johnson:2019ljv}. In black hole images, the term photon ring represents a sequence of photon subrings on the observer's screen, whereas here photon rings are unstable circular null geodesics on the equatorial plane of a rotating black hole. Beyond black hole imaging, null geodesics offer insights into quasinormal modes. Gravitational quasinormal modes describe the ringdown after a black hole merger, and their measured frequencies and damping times limit the remnant mass and spin, and enable tests of general relativity~\cite{LIGOScientific:2016lio, Isi:2019aib}. A geometric description of quasinormal modes through null geodesics is derived in the eikonal limit, which is characterized by a high angular multipole number, where wave propagation follows the geometric-optics approximation. For massless scalar modes corresponding to equatorial photon rings, the absolute magnitude of the imaginary part of the quasinormal frequency is proportional to the photon ring Lyapunov exponent~\cite{Cardoso:2008bp}. The features of photon rings and scalar quasinormal modes complementarily describe the spacetime geometry's influence on null geodesics and scalar wave propagation.

    The Penrose limit gives a plane-wave geometry near a selected null geodesic~\cite{Penrose:1976}. The coefficients of the plane-wave metric are established from the curvature components evaluated in a frame parallel transported along the geodesic, known as a parallel-propagated frame~\cite{Blau:2003dz}. These curvature components encode the tidal forces controlling deviation among the neighboring null geodesics~\cite{Pirani:1956tn}. For black hole photon rings, the Lyapunov exponents and scalar quasinormal frequencies in the eikonal limit were derived from the plane-wave geometry~\cite{Giataganas:2024hil}. This analysis was broadened to a probe string governed by the Nambu--Goto action. For the equatorial Kerr photon ring, a solution of the string equations of motion gives an induced worldsheet metric with Rindler geometry. The Unruh temperature of the worldsheet horizon sets the induced temperature, and the horizon surface gravity agrees with the photon ring Lyapunov exponent. An exponentially growing correction to the local out-of-time-order correlator (OTOC) has been derived using the gravitational eikonal approximation, with its growth exponent equal to the photon ring Lyapunov exponent~\cite{Giataganas:2026ctn}. A thermal interpretation of the eikonal quasinormal spectrum has been proposed using the induced worldsheet theory~\cite{Giataganas:2026fop}. In vacuum, the Ricci tensor null contraction is zero, whereas it need not vanish with matter present, which may modify the correspondence between the worldsheet surface gravity and the Lyapunov exponent.

    Matter changes the spacetime curvature and may affect the photon ring radius, Lyapunov exponent, and induced temperature. In weakly collisional magnetized accretion flows, the pressures parallel and perpendicular to the magnetic field may differ~\cite{Sharma:2005ez}. Pressure anisotropy has been analyzed in relativistic plasma models for black hole accretion~\cite{Chandra:2015iza} and, in general, relativistic simulations of accretion onto Kerr black holes~\cite{Foucart:2017axc}, where it can be comparable to the magnetic pressure. For static, spherically symmetric black holes, galactic matter has been represented as an anisotropic fluid, and its effects on the photon orbits, Lyapunov exponents, and gravitational radiation have been evaluated~\cite{Cardoso:2021wlq}. Despite the difficulties in constructing analytical solutions, the anticipated rotation of astrophysical black holes warrants the extension of the study to rotating geometries. Rotating black hole solutions were built numerically for anisotropic fluid halos that have vanishing radial pressure and follow a modified Hernquist density profile~\cite{Fernandes:2025osu}. Analytic bounds on the photon ring radius and Lyapunov exponent can reveal the constraints imposed by the finite asymptotic mass and physical conditions of matter. A comparison of Kerr black holes with identical asymptotic mass and rotation parameters provides a baseline for quantifying the matter effects on these quantities. The influence of anisotropic matter on the worldsheet surface gravity and induced temperature motivates testing whether the Lyapunov exponent stays bounded by the induced temperature.

    In this study, analytical bounds on the equatorial photon ring radius and Lyapunov exponent were derived for rotating black holes containing anisotropic matter. This analysis incorporates constraints set by the finite asymptotic mass and physical conditions of matter. A stationary and axisymmetric metric admitting a Killing--Yano tensor was considered, and the geometry and matter distribution were established using one mass function. For a constant equatorial equation-of-state parameter and nonnegative energy density, the photon ring radius was bounded above by the Kerr photon ring radius with the same asymptotic mass and rotation parameters for co-rotating and counter-rotating motions. The lower bounds were written in terms of the quantities evaluated at the outer horizon. The massless scalar field quasinormal modes were analyzed in the eikonal limit, and the absolute magnitude of the imaginary part of the quasinormal frequency was found to be proportional to the photon ring Lyapunov exponent. The probe string in the Penrose limit along the photon ring was then used to assess the effect of anisotropic matter on the worldsheet surface gravity and induced temperature. An induced worldsheet metric with Rindler geometry was derived, and the induced temperature was identified with the worldsheet horizon Unruh temperature. Thus, the bound $2 \pi T_\mathrm{ind} = \kappa_\mathrm{ind} \geq \lambda$ was derived, generalizing the equality for Kerr black holes to the inequality with anisotropic matter. This inequality parallels the Maldacena--Shenker--Stanford bound, written in terms of the thermal quantum system temperature~\cite{Maldacena:2015waa}, and the Hashimoto--Tanahashi bound for particle motion near black hole horizons, written in terms of the Hawking temperature~\cite{Hashimoto:2016dfz}. Our previous studies have also examined such bounds and their possible violations in black hole spacetimes~\cite{Kan:2021blg, Gwak:2022xje, Park:2023lfc, Lee:2025vih}, the frame dependence of Lyapunov exponents~\cite{Lee:2025ias}, and their applications to black hole phase transitions~\cite{Lee:2026qtn}.

    The paper is organized as follows: Sec.~\ref{sec:review} summarizes the equatorial photon ring in Kerr spacetime. Sec.~\ref{sec:Einstein_equation} develops rotating black hole spacetimes with anisotropic matter. Sec.~\ref{sec:photon_ring} derives bounds on the equatorial photon ring radius. Sec.~\ref{sec:Lambda_QNM} derives the bounds for the Lyapunov exponent and the scalar quasinormal frequencies in the eikonal limit. Sec.~\ref{sec:induced_temperature} obtains the induced temperature from the probe string worldsheet and demonstrates its relationship with the Lyapunov exponent and quasinormal frequencies. Sec.~\ref{sec:conclusion} summarizes our results and outlines avenues for future research.

\section{Review of Photon Ring in Kerr Black Hole Background} \label{sec:review}
    In this section, we review the massless particle's equatorial photon ring in a Kerr black hole. In the Boyer--Lindquist coordinates, its line element is
\begin{equation}
    \mathrm{d}s^2 = - \frac{\Gamma(r)}{\Sigma(r, \theta)} \left( \mathrm{d}t - a \sin^2\theta \, \mathrm{d}\phi \right)^2 + \frac{\Sigma(r, \theta)}{\Gamma(r)} \, \mathrm{d}r^2 + \Sigma(r, \theta) \, \mathrm{d}\theta^2 + \frac{\sin^2\theta}{\Sigma(r, \theta)} \left[ - a \, \mathrm{d}t + \left( r^2 + a^2 \right) \, \mathrm{d}\phi \right]^2,
    \label{review:metric}
\end{equation}
    where
\begin{equation}
    \Gamma(r) \equiv r^2 - 2 M_{\K} r + a^2, \qquad \Sigma(r, \theta) \equiv r^2 + a^2 \cos^2\theta,
\end{equation}
    and $M_{\K}$ and $a$ represent the Kerr black hole mass and its rotation parameter, respectively, with $a \geq 0$. The tangent vector $\mathbf{u}$ of a null particle then has the form
\begin{equation}
    \mathbf{u} = \dot{t} \partial_t + \dot{r} \partial_r + \dot{\theta} \partial_\theta + \dot{\phi} \partial_\phi,
    \label{eq:tangent_vector}
\end{equation}
    where the dot represents differentiation with respect to an affine parameter. On the equatorial plane $( \theta = \pi / 2 )$, the null geodesic $\mathbf{u}^2 = 0$ then is reduced to
\begin{equation}
    - \Gamma \left( \dot{t} - a \dot{\phi} \right)^2 + \frac{r^4}{\Gamma} \dot{r}^2 + \left[ - a \dot{t} + \left( r^2 + a^2 \right) \dot{\phi} \right]^2 = 0.
    \label{review:null_geodesic}
\end{equation}
    For a stationary, axisymmetric metric~\eqref{review:metric}, the stationary Killing vector $\boldsymbol{\xi}^{(t)} = \partial_t$ and axial Killing vector $\boldsymbol{\xi}^{(\phi)}  = \partial_\phi$ yield conserved quantities
\begin{align}
    & E \equiv - \xi^{(t)}_\mu u^\mu = \left( 1 - \frac{2 M_{\K}}{r} \right) \dot{t} + \frac{2 a M_{\K}}{r} \dot{\phi},
    \label{review:E}
    \\
    & L \equiv \xi^{(\phi)}_\mu u^\mu = - \frac{2 a M_{\K}}{r} \dot{t} + \left[ r^2 + a^2 \left( 1 + \frac{2 M_{\K}}{r} \right) \right] \dot{\phi},
    \label{review:L}
\end{align}
    where $E$ and $L$ represent the conserved energy and the particle angular momentum, respectively. Substituting these conserved quantities~\eqref{review:E}~and~\eqref{review:L} into the null geodesic~\eqref{review:null_geodesic} gives the radial equation in standard effective potential form
\begin{equation}
    \frac{\dot{r}^2}{2} + V_\mathrm{eff, \mathrm{K}}(r) = 0.
\end{equation}
    For the analysis below, we define an impact parameter $b \equiv L/E$ and write the effective potential as follows
\begin{equation}
    V_\mathrm{eff, \mathrm{K}}(r) \equiv - \frac{E^2}{2 r^3} \left[ r^3 - \left( b^2 - a^2 \right) r + 2 \left( b - a \right)^2 M_{\K} \right].
\end{equation}
    The radius of the photon ring for the Kerr black hole $r_{\PR, \mathrm{K}}$ depends on the photon ring conditions $V_\mathrm{eff, \mathrm{K}}(r_{\PR, \mathrm{K}}) = V_\mathrm{eff, \mathrm{K}}'(r_{\PR, \mathrm{K}}) = 0$. Accordingly, the impact parameter and radius obey
\begin{equation}
    b^2 - a^2 = 3 r_{\PR, \mathrm{K}}^2, \qquad b - a = \epsilon \sqrt{\frac{r_{\PR, \mathrm{K}}^3}{M_{\K}}}.
    \label{review:circular_orbit_condition}
\end{equation}
    Here, $\epsilon = \pm 1$. Substituting the second equation in~\eqref{review:circular_orbit_condition} into the first equation in~\eqref{review:circular_orbit_condition} gives the cubic equation
\begin{equation}
    \left( \sqrt{\frac{r_{\PR, \mathrm{K}}}{M_{\K}}} \right)^3 - 3 \sqrt{\frac{r_{\PR, \mathrm{K}}}{M_{\K}}} + \frac{2 \epsilon a}{M_{\K}} = 0.
    \label{review:cubic_equation}
\end{equation}
    Solving the cubic equation~\eqref{review:cubic_equation} then gives the radius
\begin{equation}
    r_{\PR, \mathrm{K}}^\pm = 4 M_{\K} \cos^2\left( \frac{1}{3} \arccos \left( \mp \frac{a}{M_{\K}} \right) \right).
\end{equation}
    Here, $r_{\PR, \mathrm{K}}^+$ and $r_{\PR, \mathrm{K}}^-$ match the co-rotating $( L > 0 )$ and counter-rotating $( L < 0 )$ photon rings around the Kerr black hole, respectively. In the nonrotating limit $a \to 0$, the two radii reduce to $3 M_{\K}$, reproducing the standard result for Schwarzschild spacetime. In the extremal limit, $a \to M_{\K}$, $r_{\PR,\mathrm{K}}^+$ and $r_{\PR,\mathrm{K}}^-$ approach $M_{\K}$ and $4 M_{\K}$, respectively. Thus, for $0 \leq a \leq M_{\K}$, the photon-ring radius lies within the range of
\begin{equation}
    M_{\K} \leq r_{\PR, \mathrm{K}}^+ \leq 3 M_{\K}, \qquad 3 M_{\K} \leq r_{\PR, \mathrm{K}}^- \leq 4 M_{\K}.
    \label{review:photon_ring_radius_Kerr}
\end{equation}

\section{Rotating Black Holes with Anisotropic Matter} \label{sec:Einstein_equation}
    To extend the Kerr solution to incorporate matter, we analyze four-dimensional stationary, axisymmetric spacetimes. The analysis below of scalar quasinormal modes uses separation of the massless Klein--Gordon equation, and a Killing tensor guarantees this separation provided it satisfies a further curvature condition. An explicit form of the curvature condition is presented in Sec. ~\ref{sec:Lambda_QNM}. Thus, the Hamilton--Jacobi equation separability ensures the existence of a Killing tensor and the associated Carter constant. Thus, we start with a general metric where the Hamilton--Jacobi equation is separable. Using $(t,r,\theta,\phi)$ coordinates, the contravariant metric has the form of ~\cite{Benenti:1979erw}
\begin{equation}
	g^{\mu\nu}(r, \theta) = \frac{1}{C_1(r, \theta)}
	\begin{pmatrix}
		C_5(r, \theta) & 0 & 0 & C_4(r, \theta)
		\\
		0 & A_2(r) & 0 & 0
		\\
		0 & 0 & B_2(\theta) & 0
		\\
		C_4(r, \theta) & 0 & 0 & C_3(r, \theta)
	\end{pmatrix},
\end{equation}
    where $C_i(r, \theta) \equiv A_i(r) + B_i(\theta)$ and $A_i$ and $B_i$ depend only on $r$ and $\theta$, respectively. The associated line element has the form
\begin{equation}
    \, \mathrm{d}s^2 = - \frac{C_1}{C_4^2 - C_3 C_5} \left( C_3 \, \mathrm{d}t^2 - 2 C_4 \, \mathrm{d}t \, \mathrm{d}\phi + C_5 \, \mathrm{d}\phi^2 \right) + \frac{C_1}{A_2} \, \mathrm{d}r^2 + \frac{C_1}{B_2} \, \mathrm{d}\theta^2.
\end{equation}
    Similarly, through the orthonormal basis $e^{\hat{a}}$, this line element is written as
\begin{equation}
    \, \mathrm{d}s^2 = - \bigl( e^{\hat{0}} \bigl)^2 + \bigl( e^{\hat{1}} \bigl)^2 + \bigl( e^{\hat{2}} \bigl)^2 + \bigl( e^{\hat{3}} \bigl)^2,
\end{equation}
    where
\begin{equation}
    e^{\hat{0}} = \frac{\sqrt{C_1}}{\sqrt{-C_5}} \, \mathrm{d}t, \qquad e^{\hat{1}} = \frac{\sqrt{C_1}}{\sqrt{A_2}} \, \mathrm{d}r, \qquad e^{\hat{2}} = \frac{\sqrt{C_1}}{\sqrt{B_2}} \, \mathrm{d}\theta, \qquad e^{\hat{3}} = \frac{\sqrt{C_1} \sqrt{-C_5}}{\sqrt{C_4^2 - C_3 C_5}} \left( -\frac{C_4}{C_5} \, \mathrm{d}t + \, \mathrm{d}\phi \right).
\end{equation}
    This metric ansatz admits a symmetric Killing tensor $K_{\mu\nu}$ that obeys the Killing equation.
\begin{equation}
    \nabla_{(\rho} K_{\mu\nu)} = 0.
\end{equation}
    Its contravariant components then are given explicitly by
\begin{equation}
	K^{\mu\nu}(r, \theta) = \frac{1}{C_1(r, \theta)}
	\begin{pmatrix}
		A_1(r) B_5(\theta) - A_5(r) B_1(\theta) & 0 & 0 & A_1(r) B_4(\theta) - A_4(r) B_1(\theta)
		\\
		0 & - A_2(r) B_1(\theta) & 0 & 0
		\\
		0 & 0 & A_1(r) B_2(\theta) & 0
		\\
		A_1(r) B_4(\theta) - A_4(r) B_1(\theta) & 0 & 0 & A_1(r) B_3(\theta) - A_3(r) B_1(\theta)
	\end{pmatrix}.
\end{equation}
    To build a parallel-propagated frame along null geodesics on the photon ring and obtain the plane-wave geometry through the Penrose limit~\cite{Penrose:1976}, the Killing tensor is assumed to arise from the Killing--Yano tensor~\cite{Yano:1952}. The Killing--Yano tensor is an antisymmetric tensor $Y_{\mu\nu}$ satisfying
\begin{equation}
    K_{\mu\nu} = Y_{\mu}{}^{\rho} Y_{\nu\rho},
\end{equation}
    and by definition, $Y_{\mu\nu}$ obeys the Killing--Yano equation
\begin{equation}
    \nabla_{(\mu} Y_{\nu)\rho} = 0.
\end{equation}
    In this orthonormal basis, the components of the Killing tensor thus are
\begin{equation}
    K^{\hat{a}\hat{b}} =
	\begin{pmatrix}
		- \frac{A_1 B_5 - A_5 B_1}{C_5} & 0 & 0 & - \frac{\left( A_4 B_5 - A_5 B_4 \right) C_1}{C_5 \sqrt{C_4^2 - C_3 C_5}}
		\\
		0 & - B_1 & 0 & 0
		\\
		0 & 0 & A_1 & 0
		\\
		- \frac{\left( A_4 B_5 - A_5 B_4 \right) C_1}{C_5 \sqrt{C_4^2 - C_3 C_5}} & 0 & 0 & -\frac{( A_1 B_3 - A_3 B_1 ) C_5^2 + ( A_1 B_5 - A_5 B_1 ) C_4^2 - 2 ( A_1 B_4 - A_4 B_1 ) C_4 C_5}{C_5 ( C_4^2 - C_3 C_5 )}
	\end{pmatrix}.
    \label{eq:Killing_tensor}
\end{equation}
    The general antisymmetric ansatz of $Y^{\hat{a}\hat{b}}$ is
    \begin{equation}
	Y^{\hat{a}\hat{b}}(r, \theta) =
	\begin{pmatrix}
		0 & Y^{\hat{0}\hat{1}}(r, \theta) & Y^{\hat{0}\hat{2}}(r, \theta) & Y^{\hat{0}\hat{3}}(r, \theta)
		\\
		-Y^{\hat{0}\hat{1}}(r, \theta) & 0 & Y^{\hat{1}\hat{2}}(r, \theta) & Y^{\hat{1}\hat{3}}(r, \theta)
		\\
		-Y^{\hat{0}\hat{2}}(r, \theta) & -Y^{\hat{1}\hat{2}}(r, \theta) & 0 & Y^{\hat{2}\hat{3}}(r, \theta)
		\\
		-Y^{\hat{0}\hat{3}}(r, \theta) & -Y^{\hat{1}\hat{3}}(r, \theta) & -Y^{\hat{2}\hat{3}}(r, \theta) & 0
	\end{pmatrix}.
    \label{eq:Killing_Yano_tensor_general}
\end{equation}
    Substitution of \eqref{eq:Killing_tensor}~and~\eqref{eq:Killing_Yano_tensor_general} into $K^{\hat{a}\hat{b}} = Y^{\hat{a}}{}_{\hat{c}} Y^{\hat{b}\hat{c}}$ yields the conditions
\begin{equation}
    A_4 B_5 \neq A_5 B_4, \qquad A_4^2 = A_3 A_5, \qquad B_4^2 = B_3 B_5.
    \label{eq:metric_conditions}
\end{equation}
    Imposing the conditions~\eqref{eq:metric_conditions}, we select, in turn, a new orthonormal basis
\begin{equation}
	e^{\hat{0}} = \frac{\sqrt{C_1} \sqrt{-A_3}}{A_3 B_4 - A_4 B_3} \left( B_3 dt - B_4 d\phi \right), \quad e^{\hat{1}} = \frac{\sqrt{C_1}}{\sqrt{A_2}} dr, \quad e^{\hat{2}} = \frac{\sqrt{C_1}}{\sqrt{B_2}} d\theta, \quad e^{\hat{3}} =  \frac{\sqrt{C_1} \sqrt{B_3}}{A_3 B_4 - A_4 B_3} \left( - A_3 dt + A_4 d\phi \right).
\end{equation}
    The Killing--Yano tensor then is reduced to
\begin{equation}
	Y^{\hat{a}\hat{b}} =
	\begin{pmatrix}
		0 & \epsilon_1 \sqrt{B_1} & 0 & 0
		\\
		-\epsilon_1 \sqrt{B_1} & 0 & 0 & 0
		\\
		0 & 0 & 0 & \epsilon_2 \sqrt{A_1}
		\\
		0 & 0 & -\epsilon_2 \sqrt{A_1} & 0
	\end{pmatrix},
\end{equation}
    where $\epsilon_i = \pm 1$. The associated components in a coordinate basis are as follows.
\begin{equation}
	Y_{\mu\nu} =
	C_1 \begin{pmatrix}
		0 & - \epsilon_1 \frac{\sqrt{-A_3} \sqrt{B_1} B_3}{\sqrt{A_2} ( A_3 B_4 - A_4 B_3 )} & \epsilon_2 \frac{\sqrt{A_1} A_3 \sqrt{B_3}}{\sqrt{B_2} ( A_3 B_4 - A_4 B_3 )} & 0
		\\
		\epsilon_1 \frac{\sqrt{-A_3} \sqrt{B_1} B_3}{\sqrt{A_2} ( A_3 B_4 - A_4 B_3 )} & 0 & 0 & - \epsilon_1 \frac{\sqrt{-A_3} \sqrt{B_1} B_4}{\sqrt{A_2} ( A_3 B_4 - A_4 B_3 )}
		\\
		- \epsilon_2 \frac{\sqrt{A_1} A_3 \sqrt{B_3}}{\sqrt{B_2} ( A_3 B_4 - A_4 B_3 )} & 0 & 0 & \epsilon_2 \frac{\sqrt{A_1} A_4 \sqrt{B_3}}{\sqrt{B_2} ( A_3 B_4 - A_4 B_3 )}
		\\
		0 & \epsilon_1 \frac{\sqrt{-A_3} \sqrt{B_1} B_4}{\sqrt{A_2} ( A_3 B_4 - A_4 B_3 )} & - \epsilon_2 \frac{\sqrt{A_1} A_4 \sqrt{B_3}}{\sqrt{B_2} ( A_3 B_4 - A_4 B_3 )} & 0
	\end{pmatrix}.
    \label{eq:Killing_Yano_tensor}
\end{equation}
    For this tensor, the Killing--Yano equation thus gives the following relationships among the metric functions
\begin{alignat}{3}
	A_3 &= - \frac{4 A_1 \left( \gamma A_1 + \delta \right)^2}{( A_1' )^2 A_2}, & \qquad A_4 &= \frac{\alpha A_1 + \beta}{\gamma A_1 + \delta} A_3, & \qquad A_5 &= \frac{A_4^2}{A_3},
	\\
	B_3 &= \frac{4 B_1 \left( \gamma B_1 - \delta \right)^2}{( B_1' )^2 B_2}, & \qquad B_4 &= \frac{\alpha B_1 - \beta}{\gamma B_1 - \delta} B_3, & \qquad B_5 &= \frac{B_4^2}{B_3},
\end{alignat}
    where primes represent derivatives relative to the corresponding function arguments and $\alpha$, $\beta$, $\gamma$, $\delta$ are four independent constants arising from two separation constants and two integration constants, with $\alpha \delta - \beta \gamma \neq 0$. We recover the Kerr black hole solution in the class of spacetimes analyzed here by choosing the Kerr metric angular sector
\begin{equation}
	B_1 = a^2 \cos^2\theta, \qquad B_2 = 1, \qquad B_3 = \frac{1}{\sin^2\theta}, \qquad B_4 = a, \qquad B_5 = a^2 \sin^2\theta,
\end{equation}
    where $a$ is the rotation parameter and $a \geq 0$. The Kerr angular functions uniquely determine the constants as
\begin{equation}
	\alpha = \pm 1, \qquad \beta = \alpha a^2, \qquad \gamma = 0, \qquad \delta = \alpha a.
\end{equation}
    The radial metric functions thus reduce to
\begin{equation}
	A_3 = - \frac{4 a^2 A_1}{( A_1' )^2 A_2}, \qquad A_4 = - \frac{4 a \left( A_1 + a^2 \right) A_1}{( A_1' )^2 A_2}, \qquad A_5 = \frac{A_4^2}{A_3}.
    \label{eq:A}
\end{equation}
    This relation~\eqref{eq:A} restricts the radial sector to two functions: $A_1(r)$ and $A_2(r)$. Redefining the radial coordinate allows us to identify $A_1$ using the new radial coordinate square, still denoted by $r$. In this coordinate system, $A_1(r) = r^2$ and $A_2(r)$ represent the independent degrees of freedom in the radial sector. We write $A_2$ as
\begin{equation}
    A_2(r) \equiv \Delta(r) = r^2 f(r) + a^2,
\end{equation}
    where $f(r)$ denotes a general radial function. The contravariant metric then takes the following form
\begin{equation}
	g^{\mu\nu} = \frac{1}{r^2 + a^2 \cos^2\theta}
	\begin{pmatrix}
		-\frac{( r^2 + a^2 )^2}{\Delta(r)} + a^2 \sin^2\theta & 0 & 0 & -\frac{a ( r^2 + a^2 )}{\Delta(r)} + a
		\\
		0 & \Delta(r) & 0 & 0
		\\
		0 & 0 & 1 & 0
		\\
		-\frac{a ( r^2 + a^2 )}{\Delta(r)} + a & 0 & 0 & -\frac{a^2}{\Delta(r)} + \frac{1}{\sin^2\theta}
	\end{pmatrix}.
    \label{eq:metric}
\end{equation}
    Accordingly, the covariant line element is expressed as 
\begin{equation}
    \mathrm{d}s^2 = - \frac{\Delta(r)}{\Sigma(r, \theta)} \left( \mathrm{d}t - a \sin^2\theta \, \mathrm{d}\phi \right)^2 + \frac{\Sigma(r, \theta)}{\Delta(r)} \, \mathrm{d}r^2 + \Sigma(r, \theta) \, \mathrm{d}\theta^2 + \frac{\sin^2\theta}{\Sigma(r, \theta)} \left[ - a \, \mathrm{d}t + \left( r^2 + a^2 \right) \, \mathrm{d}\phi \right]^2,
\end{equation}
    and the orthonormal basis then has the form
\begin{equation}
    e^{\hat{0}} = \sqrt{\frac{\Delta}{\Sigma}} \left( \mathrm{d}t -a \sin^2\theta \, \mathrm{d}\phi \right), \qquad e^{\hat{1}} = \sqrt{\frac{\Sigma}{\Delta}} \, \mathrm{d}r, \qquad e^{\hat{2}} = \sqrt{\Sigma} \, \mathrm{d}\theta, \qquad e^{\hat{3}} = \frac{\sin\theta}{\sqrt{\Sigma}} \left[ - a \, \mathrm{d}t + \left( r^2 + a^2 \right) \, \mathrm{d}\phi \right].
\end{equation}

    We consider matter with a diagonal energy--momentum tensor relative to the orthonormal basis
\begin{equation}
    T_{\hat{a}\hat{b}} = \mathrm{diag}\left( \rho(r, \theta), p_\mathrm{r}(r, \theta), p_\theta(r, \theta), p_\phi(r, \theta) \right).
\end{equation}
    The energy--momentum tensor nonzero components in the coordinate basis derived from $T_{\mu\nu} = e^{\hat{a}}_\mu e^{\hat{b}}_\nu T_{\hat{a}\hat{b}}$ are
\begin{align}
	T_{tt} &= \frac{1}{\Sigma} \left( \Delta \rho + a^2 \sin^2\theta p_\phi \right), \qquad T_{t\phi} = T_{\phi t} = -\frac{a \sin^2\theta}{\Sigma} \left[ \Delta \rho + \left( r^2 + a^2 \right) p_\phi \right],
	\\
	T_{rr} &= \frac{\Sigma}{\Delta} p_\mathrm{r}, \qquad\, T_{\theta\theta} = \Sigma p_\theta, \qquad\, T_{\phi\phi} = \frac{a^2 \sin^4\theta}{\Sigma} \left[ \Delta \rho + \frac{\left( r^2 + a^2 \right)^2}{a^2 \sin^2\theta} p_\phi \right].
\end{align}
    To represent the Einstein equations, we set
\begin{equation}
    F^\mu{}_\nu \equiv G^\mu{}_\nu - \kappa_\mathrm{g} T^\mu{}_\nu,
\end{equation}
    where $\kappa_\mathrm{g} = 8 \pi G$ in units where the speed of light is one, $G$ represents Newton's constant, and Einstein's equations become
\begin{equation}
    F^\mu{}_\nu = 0.
\end{equation}
    The two linear combinations of the Einstein equations then may be written as
\begin{align}
    & 0 = \frac{\Sigma^2}{r^3} \left( F^t{}_t + \frac{a}{r^2 + a^2} F^t{}_\phi \right) = f' - \frac{1 - f}{r} + \frac{\kappa_\mathrm{g} \Sigma^2 \rho}{r^3},
    \label{eq:einstein_equation_1}
    \\
    & 0 = \frac{\Sigma^2}{r^3} F^r{}_r = f' - \frac{1 - f}{r} - \frac{\kappa_\mathrm{g} \Sigma^2 p_\mathrm{r}}{r^3},
    \label{eq:einstein_equation_2}
\end{align}
    and the difference between Eqs.~\eqref{eq:einstein_equation_1}~and~\eqref{eq:einstein_equation_2} gives the following relationship between energy density and radial pressure
\begin{equation}
    \rho(r, \theta) = - p_\mathrm{r}(r, \theta) = \frac{g(r)}{\kappa_\mathrm{g} \Sigma(r, \theta)^2}, \qquad g(r) \equiv r^2 \left( 1 - f(r) - r f'(r) \right).
    \label{eq:energy_pressure}
\end{equation}
    We parameterize the radial metric function $f(r)$ through the mass function $M(r)$ as follows
\begin{equation}
    f(r) \equiv 1 - \frac{2 M(r)}{r}.
\end{equation}
    The mass function then is derived by integrating Eq.~\eqref{eq:einstein_equation_1} as
\begin{equation}
    M(r) = M_\mathrm{h} + \frac{\kappa_\mathrm{g}}{2} \int_{r_\mathrm{h}}^r \frac{\Sigma(s, \theta)^2}{s^2} \rho(s, \theta) \, \mathrm{d}s,
\end{equation}
    where $r_\mathrm{h}$ represents the outer event horizon radius satisfying $\Delta(r_\mathrm{h}) = 0$, the subscript $\mathrm{h}$ indicates evaluation at $r = r_\mathrm{h}$, and $M_\mathrm{h} = ( r_\mathrm{h}^2 + a^2 ) / ( 2 r_\mathrm{h} )$. Substitution of the relation in Eq. ~\eqref{eq:energy_pressure} into the remaining independent components of Einstein's equations gives
\begin{align}
    & 0 = - 2 r \Sigma F^\theta{}_\theta = g' - \frac{2 r g}{\Sigma} + 2 \kappa_\mathrm{g} r \Sigma p_\theta,
    \label{eq:einstein_equation_3}
    \\
    & 0 = - 2 r \Sigma \left( F^\phi{}_\phi - \frac{a}{r^2 + a^2} F^t{}_\phi \right) = g' - \frac{2 r g}{\Sigma} + 2 \kappa_\mathrm{g} r \Sigma p_\phi.
    \label{eq:einstein_equation_4}
\end{align}
    The difference between Eqs.~\eqref{eq:einstein_equation_3}~and~\eqref{eq:einstein_equation_4} implies the angular pressure equality and identifies their common value as the tangential pressure $p_\mathrm{t}(r, \theta)$
\begin{equation}
    p_\mathrm{t}(r, \theta) \equiv p_\theta(r, \theta) = p_\phi(r, \theta) = \rho(r, \theta) - \frac{g'(r)}{2 \kappa_\mathrm{g} r \Sigma(r, \theta)}.
\end{equation}
    Consequently, the equations, together with energy--momentum conservation, $\nabla_\rho T^\rho{}_\mu=0$, reduce to one independent equation of motion
\begin{equation}
    \rho(r, \theta) = \frac{g(r)}{\kappa_\mathrm{g} \Sigma(r, \theta)^2}.
\end{equation}
    The energy--momentum tensor in the orthonormal basis accordingly reduces to
\begin{equation}
    T_{\hat{a}\hat{b}} = \mathrm{diag}\left( \rho(r, \theta), \, - \rho(r, \theta), \, p_\mathrm{t}(r, \theta), \, p_\mathrm{t}(r, \theta) \right).
    \label{eq:energy_momentum_tensor}
\end{equation}
    Thus, specifying the radial function $g(r)$ determines the spacetime geometry and the energy--momentum tensor. The vacuum limit $g \to 0$ gives a constant mass function and reproduces Kerr spacetime, while $g = Q^2$, where $Q$ is the electric charge, yields a traceless electromagnetic energy-momentum tensor and reproduces the Kerr--Newman geometry. For a varying function $g(r)$, the source represents a general class of anisotropic matter distributions.

\section{Bounds on Equatorial Photon Ring Radius} \label{sec:photon_ring}
    Here, we investigate the rotating spacetime null geodesics governed by \eqref{eq:metric}. With $\theta$ fixed at $\pi / 2$ on the equatorial plane, the explicit argument $\theta$ is omitted. The energy density and tangential pressure then are simplified to
\begin{equation}
    \rho(r) = \frac{2 M'(r)}{\kappa_\mathrm{g} r^2}, \qquad p_\mathrm{t}(r) = - \frac{M''(r)}{\kappa_\mathrm{g} r}.
\end{equation}
    Moreover, we take a linear barotropic equation-of-state parameter $w$ on the equatorial plane
\begin{equation}
    w \equiv \frac{p_\mathrm{t}(r)}{\rho(r)} = - \frac{r M''(r)}{2 M'(r)},
    \label{eq:equation_of_state}
\end{equation}
    where the mass function approaches a finite limit $M_\infty$ at spatial infinity. For a constant equation-of-state parameter $w$, the equation~\eqref{eq:equation_of_state} integrates into
\begin{equation}
    M(r) =
    \begin{cases}
        c_1 + c_2 \log\left( \frac{r}{d} \right) & \left( w = \frac{1}{2} \right),
        \\
        c_1 + \frac{c_2}{1 - 2 w} r^{1 - 2 w} & \left( w \neq \frac{1}{2} \right),
    \end{cases}
\end{equation}
    where $c_1$ and $c_2$ denote integration constants and $d$ is a general reference length. As $r \to \infty$, the radial term diverges when $w \leq 1/2$ with nonvanishing $c_2$, but vanishes for $w > 1/2$, giving an asymptotic form
\begin{equation}
    \lim_{r \to \infty} M(r) =
    \begin{cases}
        \pm \infty & \left( w \leq \frac{1}{2} \right),
        \\
        M_\infty & \left( w > \frac{1}{2} \right).
    \end{cases}
\end{equation}
    For $w > 1/2$, the integration constant $c_1$ is identified with the asymptotic limit $M_\infty$. To ensure a well-behaved asymptotic mass value, we thus consider only $w > 1/2$.

    Energy conditions further limit the integration constant $c_2$ and equation-of-state parameter $w$. As $M'(r) = c_2 r^{-2 w}$, the energy density becomes
\begin{equation}
    \rho(r) = \frac{2c_2r^{2 ( 1 - w )}}{\kappa_\mathrm{g} \Sigma(r)^2},
\end{equation}
    and a nonnegative energy density implies $c_2 \geq 0$, with $c_2 = 0$ reproducing vacuum. For the diagonal energy-momentum tensor~\eqref{eq:energy_momentum_tensor}, the null energy condition (NEC), weak energy condition (WEC), strong energy condition (SEC), and dominant energy condition (DEC) reduce to
\begin{alignat}{3}
    \mathrm{NEC}: & \quad & \rho + p_\mathrm{t} & \geq 0, & &
    \\
    \mathrm{WEC}: & \quad & \rho & \geq 0, \quad & \rho + p_\mathrm{t} & \geq 0,
    \\
    \mathrm{SEC}: & \quad & \rho + p_\mathrm{t} & \geq 0, \quad & p_\mathrm{t} & \geq 0,
    \\
    \mathrm{DEC}: & \quad & \rho & \geq 0, \quad & \left| p_\mathrm{t} \right| & \leq \rho.
\end{alignat}
    In the equatorial plane, $p_\mathrm{t}=w\rho$, and NEC, WEC, and SEC hold for $w > 1 / 2$. In addition, the DEC imposes $w \leq 1$, giving a range
\begin{equation}
    \frac{1}{2} < w \leq 1.
\end{equation}
    For $w > 1 / 2$ and $c_2 \geq 0$, the mass function obeys
\begin{equation}
    M_\infty - M(r) = \frac{c_2}{2w-1}r^{1-2w} \geq 0,
\end{equation}
    which implies that $M_\mathrm{h} \leq M_\infty$. The horizon equation $\Delta(r_\mathrm{h}) = 0$ gives
\begin{equation}
    M(r_\mathrm{h}) = \frac{r_\mathrm{h}^2 + a^2}{2 r_\mathrm{h}} = a + \frac{(r_\mathrm{h} - a)^2}{2 r_\mathrm{h}} \geq a.
\end{equation}
    The rotation parameter consequently satisfies
\begin{equation}
    a \leq M_\infty.
\end{equation}

    With the relationship between energy density and tangential pressure fixed, we next examine the equatorial null geodesics that define the photon ring. The Hamilton--Jacobi equation for null geodesics is
\begin{equation}
    g^{\mu\nu} \partial_\mu S \partial_\nu S = 0,
    \label{eq:Hamilton_Jacobi_equation}
\end{equation}
    where $S$ denotes Hamilton's principal function and the canonical momenta are $p_\mu = \partial_\mu S$. Because the Hamilton--Jacobi equation for the metric~\eqref{eq:metric} is separable, $S$ may be written as
\begin{equation}
    S(t, r, \theta, \phi) = - E t + L \phi + S_r(r) + S_\theta(\theta).
\end{equation}
    Substituting this separated form of $S$ into the Hamilton--Jacobi equation~\eqref{eq:Hamilton_Jacobi_equation} gives the radial and angular equations
\begin{align}
    0 &= \Delta \left( S_r' \right)^2 - \frac{\left[ E \left( r^2 + a^2 \right) - a L \right]^2}{\Delta} + \mathcal{K},
    \\
    0 &= \left( S_\theta' \right)^2 + \left( \frac{L}{\sin\theta} - a E\sin\theta \right)^2 - \mathcal{K},
\end{align}
    where $\mathcal{K}$ is the separation constant equal to the conserved quantity corresponding to the Killing tensor.
\begin{equation}
    \mathcal{K} \equiv K^{\mu\nu} p_\mu p_\nu = p_\theta^2 + \left( \frac{L}{\sin\theta} - a E \sin\theta \right)^2.
\end{equation}
    The corresponding Carter constant $\mathcal{Q}$ is
\begin{equation}
    \mathcal{Q} \equiv \mathcal{K} - \left( L - a E \right)^2,
\end{equation}
    and $\mathcal{Q} = 0$ for null geodesics in an equatorial plane.

    Similarly, the radial equation governing equatorial null geodesics, $\dot{r}^2 / 2 + V_\mathrm{eff}(r) = 0$, follows from $\mathbf{u}^2 = 0$, as shown in Sec. ~\ref{sec:review}. 
\begin{equation}
    V_\mathrm{eff}(r) = - \frac{E^2}{2 r^2} \left[ r^2 - \left( b - a \right)^2 f(r) - 2 a \left( b - a \right) \right],
    \label{eq:effective_potential}
\end{equation}
    where the particle energy and angular momentum are
\begin{equation}
    E = f(r) \dot{t} + a \left( 1 - f(r) \right) \dot{\phi}, \qquad L = - a \left( 1 - f(r) \right) \dot{t} + \left[ r^2 + a^2 \left( 2 - f(r) \right) \right] \dot{\phi}.
\end{equation}
    Solving the equations for $E$ and $L$ for $\dot{t}$ and $\dot{\phi}$ gives
\begin{equation}
    \dot{t} = \frac{E}{\Delta(r)} \left[ r^2 + a \left( b - a \right) f(r) - a \left( b - 2 a \right) \right], \qquad \dot{\phi} = \frac{E}{\Delta(r)} \left[ \left( b - a \right) f(r) + a \right].
    \label{eq:tdot_phidot}
\end{equation}
    To streamline the circular-orbit conditions, we set
\begin{equation}
    N(r) \equiv \frac{r^2 f'(r)}{2} = M(r) - r M'(r).
\end{equation}
    The photon ring conditions $V_\mathrm{eff}(r_{\PR}) = 0$ and $V_\mathrm{eff}'(r_{\PR}) = 0$, with $r_{\PR}$ denoting its radius, yield
\begin{equation}
    \left( r_{\PR} - 2 M(r_{\PR}) - N(r_{\PR}) \right)^2 = \frac{4 a^2 N(r_{\PR})}{r_{\PR}}, \qquad \left( b - a \right)^2 = \frac{2 r_{\PR}}{f'(r_{\PR})} = \frac{r_{\PR}^3}{N(r_{\PR})}.
    \label{eq:circular_orbit_conditions}
\end{equation}
    For an unstable orbit, $V_\mathrm{eff}''(r_{\PR}) < 0$, requiring
\begin{equation}
    3 N(r_{\PR}) - \kappa_\mathrm{g} r_{\PR}^3 p_\mathrm{t}(r_{\PR}) > 0.
    \label{eq:unstable_orbit_condition}
\end{equation}
    For the co-rotating $(L > 0)$ and the counter-rotating $(L < 0)$ photon rings with radii $r_{\PR}^{+}$ and $r_{\PR}^{-}$, respectively, the first equation in Eq. ~\eqref{eq:circular_orbit_conditions} reduces to
\begin{align}
    0 &= r_{\PR}^+ - 2 M(r_{\PR}^+) - N(r_{\PR}^+) + 2 a \sqrt{\frac{N(r_{\PR}^+)}{r_{\PR}^+}},
    \label{eq:photon_ring_condition_1}
    \\
    0 &= r_{\PR}^- - 2 M(r_{\PR}^-) - N(r_{\PR}^-) - 2 a \sqrt{\frac{N(r_{\PR}^-)}{r_{\PR}^-}}.
    \label{eq:photon_ring_condition_2}
\end{align}
    To solve the conditions~\eqref{eq:photon_ring_condition_1}~and~\eqref{eq:photon_ring_condition_2}, we define radial functions for a general radius in the exterior $( r \geq r_\mathrm{h} )$:
\begin{equation}
    P_\pm(r) \equiv r - 2 M(r) - N(r) \pm 2 a \sqrt{\frac{N(r)}{r}}.
\end{equation}
    For Kerr spacetime, $M(r) = N(r) = M_\infty$, and each radial function $P_\pm(r)$ reduces to
\begin{equation}
    P_{\K, \pm}(r) \equiv r - 3 M_\infty \pm 2 a \sqrt{\frac{M_\infty}{r}}.
\end{equation}
    To compare with Kerr spacetime, we set
\begin{equation}
    \delta P_\pm(r) \equiv P_\pm(r) - P_{\K, \pm}(r), \qquad \delta M(r) \equiv M_\infty - M(r) \geq 0, \qquad \delta N(r) \equiv M_\infty - N(r).
\end{equation}
    From $M'(r) = c_2 r^{-2 w}$, we find $r M' = ( 2 w - 1 ) \delta M \geq 0$ for $w > 1 / 2$, implying that $M' \geq 0$, and
\begin{equation}
    \delta N = \delta M + r M' \geq 0, \qquad N' = - r M'' = 2 w M' \geq 0, \qquad \delta P_\pm = 2 \delta M + \left[ 1 \mp \frac{2 a}{\sqrt{r} \left( \sqrt{M_\infty} + \sqrt{N} \right)} \right] \delta N.
\end{equation}
    The two outer horizon conditions $\Delta(r_\mathrm{h})=0$ and $\Delta'(r_\mathrm{h}) \geq 0$ yield
\begin{equation}
    N(r) \geq N_\mathrm{h} = \frac{a^2}{r_\mathrm{h}} + \frac{\Delta'(r_\mathrm{h})}{2} \geq \frac{a^2}{r_\mathrm{h}},
    \label{eq:inequality_N}
\end{equation}
    which, used together with $M_\infty \geq N(r)$, gives
\begin{equation}
    \sqrt{r} \left( \sqrt{M_\infty} + \sqrt{N(r)} \right) \geq 2 \sqrt{r N(r)} \geq 2 \sqrt{r_\mathrm{h} N_\mathrm{h}} \geq 2 a.
\end{equation}
    Consequently, at every $r \geq r_\mathrm{h}$, $\delta P_+(r)$ obeys
\begin{equation}
    \delta P_+(r) = 2 \delta M(r) + \delta N(r) \left[ 1 - \frac{2 a}{\sqrt{r} \left( \sqrt{M_\infty} + \sqrt{N(r)} \right)} \right] \geq 0.
    \label{eq:delta_P_plus}
\end{equation}
    Because $r_{\PR, \mathrm{K}}^+ \geq M_\infty$ by the inequality~\eqref{review:photon_ring_radius_Kerr} and $0 \leq a \leq M_\infty$ for $r > r_{\PR, \mathrm{K}}^+$,
\begin{equation}
    P_{\K, +}'(r) = 1 - a \sqrt{\frac{M_\infty}{r^3}} > 0.
\end{equation}
    Combined with $P_{\K, +}(r_{\PR, \mathrm{K}}^+) = 0$, positivity of $P_{\K, +}'(r)$ implies $P_{\K, +}(r) > 0$ for $r > r_{\PR, \mathrm{K}}^+$, and inequality~\eqref{eq:delta_P_plus} precludes an exterior photon ring with $P_+(r_{\PR}^+) = 0$ with $r_{\PR}^+ > r_{\PR, \mathrm{K}}^+$. Thus, the exterior photon ring radius $r_{\PR}^+$ cannot exceed the Kerr photon ring radius $r_{\PR, \mathrm{K}}^+$.
\begin{equation}
    r_{\PR}^+ \leq r_{\PR, \mathrm{K}}^+.
\end{equation}
    For the lower bound on $r_{\PR}^+$, we analyze the following inequality
\begin{equation}
    \frac{\mathrm{d}}{\mathrm{d}r} \left( 2 M(r) + N(r) - 2 a \sqrt{\frac{N(r)}{r}} \right) = 2 M'(r) + \left( 1 - \frac{a}{\sqrt{r N(r)}} \right) N'(r) + a \sqrt{\frac{N(r)}{r^3}} \geq 0.
\end{equation}
    The photon-ring condition $P_+(r_{\PR}^+) = 0$ gives
\begin{equation}
    r_{\PR}^+ = 2 M(r_{\PR}^+) + N(r_{\PR}^+) - 2 a \sqrt{\frac{N(r_{\PR}^+)}{r_{\PR}^+}} \geq 2 M_\mathrm{h} + N_\mathrm{h} - 2 a \sqrt{\frac{N_\mathrm{h}}{r_\mathrm{h}}}.
\end{equation}
    As $2 M_\mathrm{h} = r_\mathrm{h} + a^2 / r_\mathrm{h}$, we find
\begin{equation}
    r_{\PR}^+ \geq r_\mathrm{h} + \left( \sqrt{N_\mathrm{h}} - \frac{a}{\sqrt{r_\mathrm{h}}} \right)^2.
    \label{eq:lower_bound_co_rotating}
\end{equation}
    The quantity $N_\mathrm{h}$ can be written in terms of the energy density at the outer horizon as
\begin{equation}
    N_\mathrm{h} = \frac{r_\mathrm{h}}{2} \left( 1 + \frac{a^2}{r_\mathrm{h}^2} - \kappa_\mathrm{g} r_\mathrm{h}^2 \rho_\mathrm{h} \right).
    \label{eq:N_h}
\end{equation}
    Inserting~\eqref{eq:N_h} into this lower bound~\eqref{eq:lower_bound_co_rotating} gives
\begin{equation}
    r_{\PR}^+ \geq \left[ 1 + \frac{1}{2} \left( \sqrt{1 + \frac{a^2}{r_\mathrm{h}^2} - \kappa_\mathrm{g} r_\mathrm{h}^2 \rho_\mathrm{h}} - \frac{\sqrt{2} a}{r_\mathrm{h}} \right)^2 \right] r_\mathrm{h}.
\end{equation}
    Thus, we find that the radius $r_{\PR}^+$ of any exterior co-rotating photon ring obeys
\begin{equation}
    \left[ 1 + \frac{1}{2} \left( \sqrt{1 + \frac{a^2}{r_\mathrm{h}^2} - \kappa_\mathrm{g} r_\mathrm{h}^2 \rho_\mathrm{h}} - \frac{\sqrt{2} a}{r_\mathrm{h}} \right)^2 \right] r_\mathrm{h} \leq r_{\PR}^+ \leq r_{\PR, \mathrm{K}}^+,
    \label{eq:r_plus}
\end{equation}
    and is at most the co-rotating photon ring radius of the Kerr black hole.

    For the counter-rotating radius $r_{\PR}^-$, $\delta P_-(r)$ directly gives
\begin{equation}
    \delta P_-(r) = 2 \delta M(r) + \left[ 1 + \frac{2 a}{\sqrt{r} \left( \sqrt{M_\infty} + \sqrt{N(r)} \right)} \right] \delta N(r) \geq 0.
    \label{eq:delta_P_minus}
\end{equation}
    In Kerr spacetime, $P_{\K, -}(r)$ monotonically increases.
\begin{equation}
    P_{\K,-}'(r) = 1 + a \sqrt{\frac{M_\infty}{r^3}} \geq 0,
\end{equation}
    and inequality~\eqref{eq:delta_P_minus} implies that
\begin{equation}
    r_{\PR}^- \leq r_{\PR,\mathrm{K}}^-.
    \label{eq:upper_bound_counter_rotating}
\end{equation}
    For a lower bound on $r_{\PR}^-$, the photon ring condition $P_-(r_{\PR}^-) = 0$ is written as
\begin{equation}
    r_{\PR}^- = 2 M(r_{\PR}^-) + N(r_{\PR}^-) + 2 a \sqrt{\frac{N(r_{\PR}^-)}{r_{\PR}^-}} \geq 2 M_\mathrm{h} + N_\mathrm{h} + 2 a \sqrt{\frac{N_\mathrm{h}}{r_{\PR}^-}}.
\end{equation}
    This relation, together with the inequality~\eqref{eq:inequality_N}, implies that
\begin{equation}
    r_{\PR}^- \geq r_\mathrm{h} + \frac{2 a^2}{r_\mathrm{h}} + \frac{2 a^2}{\sqrt{r_\mathrm{h} r_{\PR}^-}}.
    \label{eq:delta_P_minus_reduced}
\end{equation}
    Inequality~\eqref{eq:delta_P_minus_reduced} equivalently can be written as follows
\begin{equation}
    \left[ \sqrt{\frac{r_{\PR}^-}{r_\mathrm{h}}} + 1 \right] \left[ \left( \sqrt{\frac{r_{\PR}^-}{r_\mathrm{h}}} \right)^2 - \sqrt{\frac{r_{\PR}^-}{r_\mathrm{h}}} - 2 \left( \frac{a}{r_\mathrm{h}} \right)^2 \right] \geq 0,
\end{equation}
    yielding, for $r_{\PR}^- \geq r_\mathrm{h}$,
\begin{equation}
    \sqrt{\frac{r_{\PR}^-}{r_\mathrm{h}}} \geq \frac{1}{2} \left( 1 + \sqrt{1 + \frac{8 a^2}{r_\mathrm{h}^2} } \right).
    \label{eq:lower_bound_counter_rotating}
\end{equation}
    From the lower and upper bounds~\eqref{eq:lower_bound_counter_rotating}~and~\eqref{eq:upper_bound_counter_rotating}, it follows that the radius $r_{\PR}^-$ of every exterior counter-rotating photon ring obeys
\begin{equation}
    \frac{1}{4} \left( 1 + \sqrt{1 + \frac{8 a^2}{r_\mathrm{h}^2}} \right)^2 r_\mathrm{h} \leq r_{\PR}^- \leq r_{\PR,\mathrm{K}}^-.
    \label{eq:r_minus}
\end{equation}
    The bounds on the photon ring radii in ~\eqref{eq:r_plus} and~\eqref{eq:r_minus} combine as follows
\begin{equation}
	\left[ 1 + \frac{1}{2} \left( \sqrt{1 + \frac{a^2}{r_\mathrm{h}^2} - \kappa_\mathrm{g} r_\mathrm{h}^2 \rho_\mathrm{h}} - \frac{\sqrt{2} a}{r_\mathrm{h}} \right)^2 \right] r_\mathrm{h} \leq r_{\PR}^+ \leq r_{\PR, \mathrm{K}}^+, \qquad \frac{1}{4} \left( 1 + \sqrt{1 + \frac{8 a^2}{r_\mathrm{h}^2}} \right)^2 r_\mathrm{h} \leq r_{\PR}^- \leq r_{\PR, \mathrm{K}}^-.
\end{equation}
    For co-rotating and counter-rotating photon rings, with matter present, each radius is at most the photon ring radius of Kerr spacetime with an identical mass $M_\infty$ and rotation parameter $a$.

\section{Bounds on Lyapunov exponents and Quasinormal Modes} \label{sec:Lambda_QNM}
    The preceding section derived the bounds on the equatorial photon ring radius. Here, we examine the relationship between the equatorial photon ring Lyapunov exponent and the quasinormal mode frequency in the eikonal limit. We began by calculating the equatorial photon ring Lyapunov exponents. Measured relative to the coordinate time $t$, the Lyapunov exponent $\lambda$ quantifies the small radial displacement's exponential growth $\delta r \propto e^{\lambda t}$. Expanding the radial equation $\dot{r}^2/2 + V_\mathrm{eff} = 0$ to first order in the small radial displacement $\delta r = r - r_{\PR}^\pm$ gives~\cite{Cardoso:2008bp}
\begin{equation}
    \lambda_\pm^2 = \left. - \frac{V_{\mathrm{eff}}''}{\dot{t}^2} \right|_{r = r_{\PR}^\pm}.
    \label{eq:Lyapunov_exponent}
\end{equation}
    Using the photon-ring conditions~\eqref{eq:circular_orbit_conditions} yields the effective potential second derivative at the ring as
\begin{equation}
    V_{\mathrm{eff}}''(r_{\PR}^\pm) = \left. - E^2 \left( \frac{3}{r^2} + \frac{M''}{N} \right) \right|_{r = r_{\PR}^\pm},
    \label{eq:Vpp}
\end{equation}
    and the ratio $E / \dot{t}$ on the photon ring simplifies to
\begin{equation}
    \left. \frac{E}{\dot{t}} \right|_{r = r_{\PR}^\pm} = \left. \frac{\left( \sqrt{r N} \mp a \right) \sqrt{r N}}{r^2 \pm a \sqrt{r N}} \right|_{r = r_{\PR}^\pm},
    \label{eq:E/t}
\end{equation}
    where upper and lower signs map to $r = r_{\PR}^+$ and $r = r_{\PR}^-$, respectively. From \eqref{eq:Vpp}~and~\eqref{eq:E/t}, the Lyapunov exponent reduces to
\begin{equation}
    \lambda_\pm^2 = \frac{\left( \sqrt{r N} \mp a \right)^2}{\left( r^2 \pm a \sqrt{r N} \right)^2} \left. \left( r M'' + \frac{3 N}{r} \right) \right|_{r = r_{\PR}^\pm}.
\end{equation}
    For an unstable photon ring, the condition $V_\mathrm{eff}'' < 0$ imposes $r^2 M'' + 3 N > 0$ at the photon ring, which, for the matter with $p_\mathrm{t} = w \rho$ and $w > 1/2$, is equivalent to
\begin{equation}
    \rho(r_{\PR}^\pm) < \left. \frac{6 M}{\left( 3 + 2 w \right) \kappa_\mathrm{g} r^3} \right|_{r = r_{\PR}^\pm}.
\end{equation}
    The positive Lyapunov exponent thus is
\begin{equation}
    \lambda_\pm = \left. \frac{\sqrt{r N} \mp a}{r^2 \pm a \sqrt{r N}} \sqrt{r M'' + \frac{3 N}{r}} \right|_{r = r_{\PR}^\pm} = \left. \frac{\sqrt{r N} \mp a}{r^2 \pm a \sqrt{r N}} \sqrt{\frac{3 N - 2 w \left( M - N \right)}{r}} \right|_{r = r_{\PR}^\pm}.
\end{equation}
    As $w > 1/2$ and $M - N \geq 0$, the Lyapunov exponent obeys the upper bound:
\begin{equation}
    \lambda_\pm \leq \left. \frac{\sqrt{r N} \mp a}{r^2 \pm a \sqrt{r N}} \sqrt{\frac{4 N - M}{r}} \right|_{r = r_{\PR}^\pm}.
    \label{eq:Lyapunov_exponent_bound}
\end{equation}
    This bound becomes an equality in vacuum, for which $M = N = M_\infty$, and recovers the equatorial photon ring Lyapunov exponent in Kerr spacetime.

	Next, we consider scalar quasinormal modes. As mentioned at the beginning of Sec. ~\ref{sec:Einstein_equation}, the metric was built to allow the massless Klein--Gordon equation separation, requiring the Killing tensor to obey the further curvature condition
\begin{equation}
    0 = \nabla_\rho \left[ K, R \right]^\rho{}_\mu \equiv \nabla_\rho \left( K^{\rho\sigma} R_{\sigma\mu} - R^{\rho\sigma} K_{\sigma\mu} \right).
    \label{QNM_condition}
\end{equation}
    For the metric in~\eqref{eq:metric}, the Killing tensor takes the form
\begin{equation}
	K^{\mu\nu} = \frac{1}{r^2 + a^2 \cos^2\theta}
	\begin{pmatrix}
		a^2 r^2 \sin^2\theta + \frac{a^2 \left( r^2 + a^2 \right)^2 \cos^2\theta}{\Delta(r)} & 0 & 0 & a r^2 + \frac{a^3 \left( r^2 + a^2 \right) \cos^2\theta}{\Delta(r)}
		\\
		0 & - a^2 \Delta(r) \cos^2\theta & 0 & 0
		\\
		0 & 0 & r^2 & 0
		\\
		a r^2 + \frac{a^3 \left( r^2 + a^2 \right) \cos^2\theta}{\Delta(r)} & 0 & 0 & \frac{r^2}{\sin^2\theta} + \frac{a^4 \cos^2\theta}{\Delta(r)}
	\end{pmatrix}.
\end{equation}
    This relation obeys the condition~\eqref{QNM_condition}. A massless scalar field $\Psi$ on the background metric~\eqref{eq:metric} satisfies the Klein--Gordon equation
\begin{equation}
    \Box \Psi \equiv \frac{1}{\sqrt{-g}} \partial_\mu \left( \sqrt{-g} \, g^{\mu\nu} \partial_\nu \Psi \right) = 0.
\label{eq:Klein_Gordon_equation}
\end{equation}
    The Klein--Gordon equation is identical in form to the spin-zero Teukolsky equation~\cite{Teukolsky:1972my}, with the constant mass $M$ replaced by the mass function $M(r)$. As the separability condition~\eqref{QNM_condition} holds, the scalar field $\Psi$ admits a separated form
\begin{equation}
    \Psi(t, r, \theta, \phi) = e^{- i \omega t} e^{i m \phi} \frac{\mathcal{R}(r)}{r} \Theta(\theta).
\end{equation}
    Here, $\omega = \omega_\mathrm{R} - i \omega_\mathrm{I}$ is a complex frequency and $m$ is an azimuthal number. Substituting into the Klein--Gordon equation~\eqref{eq:Klein_Gordon_equation} reduces it to
\begin{equation}
    0 = \Delta \frac{\mathcal{R}''}{\mathcal{R}} + \left( \Delta' - \frac{2 \Delta}{r} \right) \frac{\mathcal{R}'}{\mathcal{R}} -\frac{\Delta'}{r} + \frac{2 \Delta}{r^2} + \frac{\left[ \omega \left( r^2 + a^2 \right) - m a \right]^2}{\Delta} + \frac{\Theta''}{\Theta} +\cot\theta \frac{\Theta'}{\Theta} - \left( a \omega \sin\theta - \frac{m}{\sin\theta} \right)^2.
\end{equation}
    By introducing the separation constants $\Lambda_{\ell m}$ and $A_{\ell m}$, we find, in turn, the following radial and angular equations, respectively
\begin{align}
    0 &= \mathcal{R}'' + \left( \frac{\Delta'}{\Delta} - \frac{2}{r} \right) \mathcal{R}' + \left\{ \frac{2}{r^2} - \frac{\Delta'}{r \Delta} + \frac{\left[ \left( r^2 + a^2 \right) \omega - m a \right]^2}{\Delta^2} - \frac{\Lambda_{\ell m}}{\Delta} \right\} \mathcal{R},
    \\
    0 &= \frac{1}{\sin\theta} \frac{\mathrm{d}}{\mathrm{d}\theta} \left( \sin\theta \frac{\mathrm{d}\Theta}{\mathrm{d}\theta} \right) + \left[ A_{\ell m} + \omega^2 a^2 \cos^2\theta - \frac{m^2}{\sin^2\theta} \right]\Theta,
\end{align}
    with
\begin{equation}
    \Lambda_{\ell m} = A_{\ell m} - m^2 + \left( m - \omega a \right)^2.
    \label{eq:separation_constants}
\end{equation}
    To derive a Schr\"odinger-type equation, we define a new coordinate $y$ and a rescaled radial function $Z$,
\begin{equation}
    \frac{\mathrm{d}y}{dr} \equiv \frac{r^2+a^2}{\Delta(r)}, \qquad Z(r) \equiv \frac{\sqrt{r^2 + a^2}}{r} \mathcal{R}(r).
    \label{eq:coordinate_transformation}
\end{equation}
    Coordinate $y$ tends to $-\infty$ at the outer horizon and $+\infty$ at spatial infinity
\begin{equation}
    \lim_{r \to r_\mathrm{h}} y(r) = - \infty, \qquad \lim_{r \to \infty} y(r) = + \infty.
\end{equation}
    The coordinate transformation and radial rescaling~\eqref{eq:coordinate_transformation} remove the first derivative term relative to $y$, giving the Schr\"odinger-type equation
\begin{equation}
    \frac{\mathrm{d}^2Z}{\mathrm{d}y^2} + \left( \omega^2 - U \right) Z = 0.
    \label{eq:radial_equation}
\end{equation}
    Here, the effective potential for the scalar field $U$ is
\begin{equation}
    U(r) = - \frac{3 r^2 \Delta(r)^2}{\left( r^2 + a^2 \right)^4} + \frac{\left( r \Delta'(r) + \Delta(r) \right) \Delta(r)}{\left( r^2 + a^2 \right)^3} + \frac{\Lambda_{\ell m} \Delta(r) - m^2 a^2}{\left( r^2 + a^2 \right)^2} + \frac{2 \omega m a}{r^2 + a^2}.
\end{equation}
    At the leading order in the eikonal limit, the effective potential exhibits a maximum at $y = y_{\PR}$, which matches the photon ring radius $r = r_{\PR}$, where
\begin{equation}
    U_{\PR} \equiv U(y_{\PR}), \qquad U_{\PR}' \equiv \left. \frac{\partial U}{\partial y} \right|_{y = y_{\PR}} = 0, \qquad U_{\PR}'' \equiv \left. \frac{\partial^2 U}{\partial y^2} \right|_{y = y_{\PR}} < 0.
    \label{eq:effective_potential_WKB}
\end{equation}
    Expanding the effective potential to quadratic order in the vicinity of $y = y_{\PR}$ yields
\begin{equation}
    U \simeq U_{\PR} + \frac{1}{2} U_{\PR}'' \left( y - y_{\PR} \right)^2.
\end{equation}
    The radial equation~\eqref{eq:radial_equation} reduces to
\begin{equation}
    \frac{\mathrm{d}^2Z}{\mathrm{d}y^2} + \left[ \omega^2 - U_{\PR} - \frac{1}{2} U_{\PR}'' \left( y - y_{\PR} \right)^2 \right] Z = 0.
    \label{eq:radial_equation_expanded}
\end{equation}
    To cast the radial equation as a parabolic cylinder equation, we set the complex variable $z$ as
\begin{equation}
    z = e^{- i \pi / 4} \left( - 2 U_{\PR}'' \right)^{1/4} \left( y - y_{\PR} \right).
\end{equation}
    Here, the radial equation~\eqref{eq:radial_equation_expanded} becomes a parabolic cylinder form
\begin{equation}
    \frac{\mathrm{d}^2Z}{\mathrm{d}z^2} + \left( \nu + \frac{1}{2} - \frac{z^2}{4} \right) Z = 0, \qquad \nu + \frac{1}{2} = i \frac{\omega^2 - U_{\PR}}{\sqrt{- 2 U_{\PR}''}}.
\end{equation}
    Quasinormal modes satisfy purely ingoing boundary conditions at the event horizon and purely outgoing boundary conditions at spatial infinity. For time dependence $e^{-i\omega t}$, the radial boundary conditions thus are
\begin{equation}
    Z(y) \sim
    \begin{cases}
        e^{- i \left[ \omega - \left( m a \right) / \left( r_{\mathrm h}^2 + a^2 \right) \right] y}, & y \to - \infty,
        \\
        e^{+ i \omega y}, & y \to + \infty.
    \end{cases}
    \label{eq:boundary_conditions}
\end{equation}
    Matching the parabolic cylinder solution to the Wentzel--Kramers--Brillouin (WKB) solutions and imposing the boundary conditions~\eqref{eq:boundary_conditions} gives the first-order WKB condition~\cite{Schutz:1985, Iyer:1986np},
\begin{equation}
    i \frac{\omega^2 - U_{\PR}}{\sqrt{- 2 U_{\PR}''}} \simeq n + \frac{1}{2}, \qquad n = 0, 1, 2, \cdots,
\end{equation}
    where $n$ is the overtone number, with $n = 0$ corresponding to the fundamental mode. Equivalently, the frequencies of the quasinormal modes obey
\begin{equation}
    \omega^2 = U_{\PR} - i \left( n + \frac{1}{2} \right) \sqrt{- 2 U_{\PR}''} \, .
    \label{eq:WKB_condition}
\end{equation}
    In the eikonal limit, $\ell \gg 1$, where $\ell$ is the angular multipole index, and the equatorial photon rings map to modes with $|m| = \ell$. The separation constants~\eqref{eq:separation_constants} have the asymptotic forms~\cite{Dolan:2010wr}
\begin{equation}
    A_{\ell m} = \ell^2 + O(\ell), \qquad \Lambda_{\ell m} = \left( m - a \omega \right)^2 + O(\ell).
\end{equation}
    The real and imaginary parts of the frequency scale as $\omega_\mathrm{R} = O(\ell)$ and $\omega_\mathrm{I} = O(1)$. The eikonal limit allows a geometrical optics description~\cite{Yang:2012he}, where $\omega_\mathrm{R} = E$ and $m = L$. Because $\omega_\mathrm{I} / \omega_\mathrm{R} = O(\ell^{-1})$, the expansion in the imaginary part of the frequency yields
\begin{align}
    \omega^2 - U_{\PR} &\simeq \omega_\mathrm{R}^2 - 2 i \omega_\mathrm{R} \omega_\mathrm{I} - U(r_{\PR}, \omega_\mathrm{R}) + i \omega_\mathrm{I} \left. \frac{\partial U}{\partial \omega} \right|_{\{ r = r_{\PR}, \, \omega = \omega_\mathrm{R} \}}
    \\
    &= \omega_\mathrm{R}^2 - U(r_{\PR}, \omega_\mathrm{R}) - i \omega_\mathrm{I} \left( 2 \omega_\mathrm{R} - \left. \frac{\partial U}{\partial \omega} \right|_{\{ r = r_{\PR}, \, \omega = \omega_\mathrm{R} \}} \right).
    \label{eq:eikonal_limit}
\end{align}
    In the eikonal limit, the effective potential for the scalar field~\eqref{eq:effective_potential_WKB} can be written in terms of the effective potential for equatorial null geodesics~\eqref{eq:effective_potential} as
\begin{equation}
    U(r_{\PR}, \omega_\mathrm{R}) \simeq \frac{\left( m - \omega_\mathrm{R} a \right)^2 \Delta(r_{\PR}) - m^2 a^2}{\left( r_{\PR}^2 + a^2 \right)^2} + \frac{2 \omega_\mathrm{R} m a}{r_{\PR}^2 + a^2} = \omega_\mathrm{R}^2 + \frac{2 r_{\PR}^4}{\left( r_{\PR}^2 + a^2 \right)^2} V_\mathrm{eff}(r_{\PR}).
\end{equation}
    Substituting the expansion~\eqref{eq:eikonal_limit} into the first-order WKB condition~\eqref{eq:WKB_condition} then gives
\begin{equation}
    \omega_\mathrm{R}^2 = U(r_{\PR}, \omega_\mathrm{R}), \qquad \omega_\mathrm{I} = \left( n + \frac{1}{2} \right) \left. \frac{\sqrt{- 2 U_{\PR}''}}{2 \omega_\mathrm{R} - \frac{\partial U}{\partial \omega}} \right|_{\{ r = r_{\PR}, \, \omega = \omega_\mathrm{R} \}}.
    \label{WKB_conditions_eikonal_limit}
\end{equation}
    At this maximum, $r = r_{\PR}$, we find
\begin{equation}
    U_{\PR}'' = \left. \frac{\partial^2 U}{\partial y^2} \right|_{\{y = y_{\PR}, \, \omega = \omega_\mathrm{R}\}} = \frac{\Delta(r_{\PR})^2}{\left( r_{\PR}^2 + a^2 \right)^2} \left. \frac{\partial^2 U}{\partial r^2} \right|_{\{r = r_{\PR}, \, \omega = \omega_\mathrm{R}\}} = \frac{2 r_{\PR}^4 \Delta(r_{\PR})^2}{\left( r_{\PR}^2 + a^2 \right)^4} V_\mathrm{eff}''(r_{\PR}).
\end{equation}
    This relation directly leads to
\begin{equation}
    \sqrt{- 2 U_{\PR}''} = \frac{2 r_{\PR}^2 \Delta(r_{\PR})}{\left( r_{\PR}^2 + a^2 \right)^2} \sqrt{-V_\mathrm{eff}''(r_{\PR})} \, .
\end{equation}
    The denominator of the imaginary part of the frequency~\eqref{WKB_conditions_eikonal_limit} is
\begin{equation}
    2 \omega_\mathrm{R} - \left. \frac{\partial U}{\partial \omega} \right|_{\{ r = r_{\PR}, \, \omega = \omega_\mathrm{R} \}} = \frac{2 r_{\PR}^2}{\left( r_{\PR}^2 + a^2 \right)^2} \left[ \omega_\mathrm{R} \left( r_{\PR}^2 + 2 a^2 \right) + a \left( m - a \omega_\mathrm{R} \right) f(r_{\PR}) - m a \right].
\end{equation}
    By combining the expression for $\dot{t}$ in~\eqref{eq:tdot_phidot} with the Lyapunov exponent definition in~\eqref{eq:Lyapunov_exponent}, we find
\begin{equation}
    \left. \frac{\sqrt{- 2 U_{\PR}''}}{2 \omega_\mathrm{R} - \frac{\partial U}{\partial \omega}} \right|_{\{ r = r_{\PR}, \, \omega = \omega_\mathrm{R} \}} = \frac{\Delta(r_{\PR}) \sqrt{-V_\mathrm{eff}''(r_{\PR})}}{E \left[ r_{\PR}^2 + a \left( b - a \right) f(r_{\PR}) - a \left( b - 2 a \right) \right]} = \left. \frac{\sqrt{-V_\mathrm{eff}''}}{\dot{t}} \right|_{r = r_{\PR}} = \lambda.
\end{equation}
    Hence, in the eikonal limit, the absolute magnitude of the imaginary part of the quasinormal frequency depends on the equatorial photon ring Lyapunov exponent.
\begin{equation}
    \omega_\mathrm{I} = \left( n + \frac{1}{2} \right) \lambda.
\end{equation}
    This governs the decay rate of the scalar perturbations. At each fixed overtone number $n$, the upper bound on the Lyapunov exponent~\eqref{eq:Lyapunov_exponent_bound} gives an upper bound on the absolute magnitude of the imaginary part of the quasinormal frequency in the eikonal limit.
\begin{equation}
    \omega_\mathrm{I} \leq \left( n + \frac{1}{2} \right) \left. \frac{\sqrt{r N} \mp a}{r^2 \pm a \sqrt{r N}} \sqrt{\frac{4 N - M}{r}} \right|_{r = r_{\PR}^\pm}.
    \label{eq:omega_I_bound}
\end{equation}

\section{Bounds on Induced Temperature} \label{sec:induced_temperature}
    In this section, we investigate the bounds on the induced temperature, which is interpreted as the Unruh temperature for a probe string, and examine the relation between the induced temperature and the Lyapunov exponent derived in Sec. ~\ref{sec:Lambda_QNM}. The construction of plane-wave geometry imposes a parallel-propagated frame along the null geodesic~\cite{Marck:1983a, Marck:1983b, Kamran:1986, Kubiznak:2008zs}. For this construction, we recall that the Killing--Yano tensor~\eqref{eq:Killing_Yano_tensor} takes the following form
\begin{equation}
	Y_{\mu\nu} =
	\begin{pmatrix}
		0 & - \epsilon_1 a \cos\theta & - \epsilon_2 a r \sin\theta & 0
		\\
		\epsilon_1 a \cos\theta & 0 & 0 & - \epsilon_1 a^2 \sin^2\theta \cos\theta
		\\
		\epsilon_2 a r \sin\theta & 0 & 0 & - \epsilon_2 \left( r^2 + a^2 \right) r \sin\theta
		\\
		0 & \epsilon_1 a^2 \sin^2\theta \cos\theta & \epsilon_2 \left( r^2 + a^2 \right) r \sin\theta & 0
	\end{pmatrix},
\end{equation}
    and, correspondingly, the dual tensor is
\begin{equation}
	(\star Y)_{\mu\nu} \equiv \frac{1}{2} \epsilon_{\mu\nu\rho\sigma} Y^{\rho\sigma} =
	\begin{pmatrix}
		0 & - \epsilon_2 r & \epsilon_1 a^2 \sin\theta \cos\theta & 0
		\\
		\epsilon_2 r & 0 & 0 & - \epsilon_2 a r \sin^2\theta
		\\
		- \epsilon_1 a^2 \sin\theta \cos\theta & 0 & 0 & \epsilon_1 a \left( r^2 + a^2 \right) \sin\theta \cos\theta
		\\
		0 & \epsilon_2 a r \sin^2\theta & - \epsilon_1 a \left( r^2 + a^2 \right) \sin\theta \cos\theta & 0
	\end{pmatrix},
\end{equation}
    whewe $\epsilon_{\mu\nu\rho\sigma}$ is the Levi--Civita tensor. We next construct a parallel-propagated frame along an equatorial null geodesic in a photon ring. One of the frame null vectors is given by the tangent vector $\mathbf{u}$~\eqref{eq:tangent_vector}, satisfying
\begin{equation}
    u^\mu = \frac{\mathrm{d}x^\mu}{\mathrm{d}k}, \qquad u^\mu u_\mu = 0, \qquad u^\rho \nabla_\rho u^\mu = 0,
    \label{eq:u}
\end{equation}
    where $k$ is an affine parameter of the geodesic. The Killing--Yano tensor implies that the contraction $u_\nu Y^{\nu\mu}$ is propagated parallel to the null geodesic
\begin{equation}
    u^\rho \nabla_\rho \left( u^\nu Y_{\nu\mu} \right) = 0.
\end{equation}
    This allows a normalized parallel-transported vector to be defined by
\begin{equation}
    e_{(1)}^\mu = \frac{1}{\sqrt{K}} u_\nu Y^{\nu\mu},
    \label{eq:e_1}
\end{equation}
    where $K \equiv K_{\rho\sigma} u^\rho u^\sigma$. A second normalized parallel-transported vector is built from the dual Killing--Yano tensor after adding a term proportional to $u^\mu$
\begin{equation}
    e_{(2)}^\mu = \frac{1}{\sqrt{H}} \left( u_\nu (\star Y)^{\nu\mu} + k u_\nu \zeta^\nu u^\mu\right),
    \label{eq:e_2}
\end{equation}
    with normalization constant $H$ then defined by
\begin{equation}
    H \equiv H_{\mu\nu}u^\mu u^\nu, \qquad H_{\mu\nu} \equiv (\star Y)_\mu{}^\rho (\star Y)_{\nu\rho}.
\end{equation}
    Here, with the choice $\epsilon \equiv \epsilon_1 = - \epsilon_2$, the stationary Killing vector $\zeta^\mu$ has the form
\begin{equation}
    \zeta^\mu \equiv \frac{1}{3}\nabla_\nu(\star Y)^{\nu\mu} = (\epsilon, \, 0, \, 0, \, 0).
\end{equation}
    The auxiliary null vector \(n^\mu\) closes the frame.
\begin{equation}
	n^\mu = \frac{1}{\sqrt{H}} e^{(2)}_\rho (\star Y)^{\rho\mu} + \frac{1}{2 H^2} \left[ H_{\alpha\beta} H^{\beta}{}_{\gamma} u^\alpha u^\gamma + k^2 H \left( \zeta_\alpha u^\alpha \right)^2 \right] u^\mu.
    \label{eq:n}
\end{equation} 
    The derived parallel-transported frame obeys the condition
\begin{equation}
    u^2 = n^2 = 0, \qquad u \cdot n = -1, \qquad e_{(i)} \cdot e_{(j)} = \delta_{ij}, \qquad u \cdot e_{(i)} = n \cdot e_{(j)} = 0.
\end{equation}
   The metric can be written in terms of the parallel-transported frame as
\begin{equation}
    g^{\mu\nu} = - u^\mu n^\nu - n^\mu u^\nu + \delta^{ij} e_{(i)}^\mu e_{(j)}^\nu.
\end{equation}
    The Penrose limit about the equatorial photon ring gives the plane-wave metric using Brinkmann coordinates $(x^+, \, x^-, \, x^1, \, x^2)$:
\begin{equation}
    \mathrm{d}s^2 = A_{ij} x^i x^j ( \mathrm{d}x^+)^2 + 2 \mathrm{d}x^+ \mathrm{d}x^- + ( \mathrm{d}x^1 )^2 + ( \mathrm{d}x^2 )^2, \qquad i, \, j = 1, \, 2,
    \label{eq:plane_wave_geometry}
\end{equation}
    where
\begin{equation}
    A_{ij} = \left. - R_{\mu\nu\rho\sigma} u^\mu e_{(i)}^\nu u^\rho e_{(j)}^\sigma \right|_{r = r_{\PR}}.
\end{equation}
    In the parallel-transported frame specified by~\eqref{eq:u},~\eqref{eq:e_1},~\eqref{eq:e_2},~and~\eqref{eq:n}, the nonzero components of $A_{ij}$ then are
\begin{equation}
    A_{11} = \left. V_{\mathrm{eff}}'' - \frac{\kappa_\mathrm{g} E^2 r \left( \rho + p_\mathrm{t} \right)}{N} \right|_{r = r_{\PR}}, \qquad A_{22} = \left. - V_{\mathrm{eff}}'' \right|_{r = r_{\PR}}, \qquad A_{12} = A_{21} = 0,
    \label{eq:A_matrix}
\end{equation}
    and in vacuum, $A_{11} = - A_{22}$.

    To derive the induced temperature $T_\mathrm{ind}$, defined as the worldsheet horizon Unruh temperature, we analyze a probe string embedded into the plane-wave geometry~\eqref{eq:plane_wave_geometry} derived from the Penrose limit along the equatorial photon ring. We use the Nambu--Goto action~\cite{Goto:1971ce} for the probe string.
\begin{equation}
    S_\mathrm{NG} = - T_\mathrm{s} \int \mathrm{d}^2 \chi \sqrt{- \det \mathcal{G}}, \qquad \mathcal{G}_{ab} = g_{\mu\nu} \frac{\partial X^\mu}{\partial \chi^a} \frac{\partial X^\nu}{\partial \chi^b},
\end{equation}
     where $T_\mathrm{s}$ is the string tension, $\chi^a = (\tau, \sigma)$ are worldsheet coordinates, and $\mathcal{G}_{ab}$ is the metric induced by the embedding
\begin{equation}
    X^\mu (\tau, \sigma) = \left( x^+(\tau, \sigma), \, x^-(\tau, \sigma), \, x^1(\tau, \sigma), \, x^2(\tau, \sigma) \right).
\end{equation}
    We choose the null coordinate $x^+$ as the affine parameter $k$ along the photon ring and choose the static gauges $t = \tau$ and $x^1 = \sigma$. Because $\dot{t}$ in~\eqref{eq:tdot_phidot} remains constant on the photon ring, the embedding is reduced to
\begin{equation}
    X^\mu(\tau, \sigma) = \left( \frac{\tau}{\dot{t}_{\PR}}, \, x^-(\tau, \sigma), \, \sigma, \, x^2(\tau, \sigma) \right), \qquad \dot{t}_{\PR} \equiv \left. \frac{\mathrm{d}t}{\mathrm{d}k} \right|_{\PR}.
\end{equation}
    The induced worldsheet metric is
\begin{equation}
    \mathrm{d}s_\mathrm{ws}^2 = \left[ \frac{A_{11}}{\dot{t}_{\PR}^2} \sigma^2 + \frac{A_{22}}{\dot{t}_{\PR}^2} ( x^2 )^2 + ( \partial_\tau x^2 )^2 + \frac{2 \partial_\tau x^-}{\dot{t}_{\PR}} \right] \mathrm{d}\tau^2 + 2 \left( \partial_\tau x^2 \partial_\sigma x^2 + \frac{\partial_\sigma x^-}{\dot{t}_{\PR}} \right) \mathrm{d}\tau \mathrm{d}\sigma + \left[ 1 + ( \partial_\sigma x^2 )^2 \right] \mathrm{d}\sigma^2.
\end{equation}
    The Nambu--Goto Lagrangian density is
\begin{equation}
    \mathcal{L}_\mathrm{NG} = - T_\mathrm{s} \left\{ \left( \partial_\tau x^2 \partial_\sigma x^2 + \frac{\partial_\sigma x^-}{\dot{t}_{\PR}} \right)^2 - \left[ \frac{A_{11}}{\dot{t}_{\PR}^2} \sigma^2 + \frac{A_{22}}{\dot{t}_{\PR}^2} ( x^2 )^2 + ( \partial_\tau x^2 )^2 + \frac{2 \partial_\tau x^-}{\dot{t}_{\PR}} \right] \left[ 1 + ( \partial_\sigma x^2 )^2 \right] \right\}^{1/2},
\end{equation}
    and the variation relative to the remaining dynamical fields $x^-$ and $x^2$ yields the equations of motion
\begin{equation}
    \frac{\partial \mathcal{L}_\mathrm{NG}}{\partial X^\mu} - \partial_\tau \left[ \frac{\partial \mathcal{L}_\mathrm{NG}}{\partial ( \partial_\tau X^\mu )} \right] - \partial_\sigma \left[ \frac{\partial \mathcal{L}_\mathrm{NG}}{\partial ( \partial_\sigma X^\mu )} \right] = 0.
\end{equation}
    The equations of motion then hold as follows
\begin{equation}
    x^-(\tau, \sigma) = 0, \qquad x^2(\tau, \sigma) = 0.
\end{equation}
    The induced worldsheet metric consequently reduces to Rindler geometry,
\begin{equation}
    \mathrm{d}s_\mathrm{ws}^2 = \frac{A_{11}}{\dot{t}_{\PR}^2} \sigma^2 \mathrm{d}\tau^2 + \mathrm{d}\sigma^2 \equiv - \kappa_\mathrm{ind}^2 \sigma^2 \mathrm{d}\tau^2 + \mathrm{d}\sigma^2.
\end{equation}
    The surface gravity $\kappa_\mathrm{ind}$ on the worldsheet Rindler horizon, $\sigma = 0$, is
\begin{equation}
    \kappa_\mathrm{ind}^2 = - \frac{A_{11}}{\dot{t}_{\PR}^2} = \left. \left[ - \frac{V_\mathrm{eff}''}{\dot{t}^2} + \frac{\kappa_\mathrm{g} E^2 r \left( \rho + p_\mathrm{t} \right)}{N \dot{t}^2} \right] \right|_{r = r_{\PR}} = \left( \left. 1 + \frac{\kappa_\mathrm{g} r^3 \left( \rho + p_\mathrm{t} \right)}{3 N - \kappa_\mathrm{g} r^3 p_\mathrm{t}} \right|_{r = r_{\PR}} \right) \lambda^2.
    \label{eq:difference}
\end{equation}
    The components $A_{ij}$ encode the tidal forces on neighboring null geodesics, as described by the geodesic deviation equation
\begin{equation}
    \frac{\mathrm{D}^2 V^i}{\mathrm{d}k^2} = A^{i}{}_{j} V^j,
\end{equation}
    where $V^i$ is the deviation vector in the parallel-transported frame and $\mathrm{D}/\mathrm{d}k = u^\mu \nabla_\mu$ is covariant differentiation along the photon ring. The trace of $A_{ij}$ depends on the Ricci tensor null contraction
\begin{equation}
    \operatorname{tr} A = - R_{\mu\nu} u^\mu u^\nu.
\end{equation}
    From the relations~\eqref{eq:A_matrix}~and~\eqref{eq:difference}, we find
\begin{equation}
    \frac{\kappa_\mathrm{ind}^2 - \lambda^2}{\lambda^2} = - \frac{\operatorname{tr} A}{\dot{t}_{\PR}^2 \lambda^2} = \left. \frac{R_{\mu\nu} u^\mu u^\nu}{\dot{t}^2 \lambda^2} \right|_{r = r_{\PR}} = \left. \frac{\kappa_\mathrm{g} r^3 \left( \rho + p_\mathrm{t} \right)}{3 N - \kappa_\mathrm{g} r^3 p_\mathrm{t}} \right|_{r = r_{\PR}}.
\end{equation}
    The normalized difference $(\kappa_\mathrm{ind}^2 - \lambda^2) / \lambda^2$ quantifies matter's contribution to the spacetime curvature sampled by null geodesics on the photon ring. The worldsheet surface gravity thermally characterizes the Rindler horizon through the induced temperature, $T_\mathrm{ind}$, which is defined as the Unruh temperature
\begin{equation}
    T_\mathrm{ind} \equiv \frac{\kappa_\mathrm{ind}}{2 \pi} = \left. \frac{\lambda}{2 \pi} \sqrt{1 + \frac{\kappa_\mathrm{g} r^3 \left( \rho + p_\mathrm{t} \right)}{3 N - \kappa_\mathrm{g} r^3 p_\mathrm{t}}} \right|_{r = r_{\PR}}.
\end{equation}
    For an equation-of-state parameter $w > 1/2$, the condition for an unstable photon ring~\eqref{eq:unstable_orbit_condition} implies the bound
\begin{equation}
    2 \pi T_\mathrm{ind} = \kappa_\mathrm{ind} \geq \lambda.
\end{equation}
    Thus, the induced temperature gives an upper bound on the Lyapunov exponent. In vacuum, the bound becomes an equality at $2 \pi T_\mathrm{ind} = \kappa_\mathrm{ind} = \lambda$, which reproduces the result in Ref. ~\cite{Giataganas:2026ctn}. Furthermore, the bound on the absolute magnitude of the imaginary part of the quasinormal frequency~\eqref{eq:omega_I_bound} can be written in terms of the induced temperature $T_\mathrm{ind}$ as follows:
\begin{equation}
    \omega_\mathrm{I} \leq \left. 2 \pi T_\mathrm{ind} \left( n + \frac{1}{2} \right) \sqrt{\frac{4 N - M}{N + 2M}} \right|_{r = r_{\PR}}.
    \label{eq:omega_I_bound_induced_temperature}
\end{equation}
    This holds in the eikonal limit at each fixed overtone number $n$. For the fundamental mode $(n = 0)$, the lower bound on the damping time $t_\mathrm{d}$ is
\begin{equation}
    t_\mathrm{d} \equiv \frac{1}{\omega_\mathrm{I}(n = 0)} \geq \left. \frac{1}{\pi T_\mathrm{ind}} \sqrt{\frac{N + 2M}{4 N - M}} \right|_{r = r_{\PR}}.
    \label{eq:damping_time_bound_induced_temperature}
\end{equation}
    For $M = N$, which corresponds to vacuum, the two bounds~\eqref{eq:omega_I_bound_induced_temperature}~and~\eqref{eq:damping_time_bound_induced_temperature} become equalities, giving $\omega_\mathrm{I} = 2 \pi T_\mathrm{ind} ( n + 1/2 )$ and $t_\mathrm{d} = 1 / ( \pi T_\mathrm{ind} )$. For a positive energy density $M > N$, matter tightens the bounds to $\omega_\mathrm{I} < 2 \pi T_\mathrm{ind} ( n + 1/2 )$ and $t_\mathrm{d} > 1 / ( \pi T_\mathrm{ind} )$. Thus, the induced temperature limits the magnitude of the imaginary part of the quasinormal frequency and the damping time, with corrections set by the matter distribution at the photon ring.

\section{Conclusion} \label{sec:conclusion}    
    Here, analytical bounds were obtained for the equatorial photon ring radius, Lyapunov exponent, and magnitude of the imaginary part of the quasinormal mode frequency of a massless scalar field in rotating black hole spacetimes with anisotropic matter. The photon ring radius and Lyapunov exponent were established from the null geodesics, and the quasinormal mode frequency was derived from the massless Klein--Gordon equation in the eikonal limit. The Penrose limit along the photon ring produced a plane-wave geometry where the induced worldsheet metric of the probe string was reduced to the Rindler metric. The surface gravity of the worldsheet horizon defines the Unruh temperature, which is referred to as the induced temperature. The worldsheet surface gravity provides an upper bound on the photon ring Lyapunov exponent: $2 \pi T_\mathrm{ind} = \kappa_\mathrm{ind} \geq \lambda$.

    To construct a spacetime geometry where the massless Klein--Gordon equation is separable, a Killing tensor is required to obey the curvature condition that relates the Killing tensor to the Ricci tensor. Thus, the construction started with a stationary and axisymmetric metric where the Hamilton--Jacobi equation was separable, which ensures the existence of a Killing tensor. The Killing tensor was assumed to be generated by a Killing--Yano tensor to build a parallel-propagated frame along the photon ring and obtain the plane-wave geometry at the Penrose limit. The metric functions separate into radial and angular sectors, and the Killing--Yano equations were used to constrain the radial and angular functions through four independent constants. The angular sector was set equal to the Kerr metric angular sector to recover the Kerr solution in vacuum while fixing the four constants. Redefinition of the radial coordinates reduces the remaining radial freedom to a single mass function. The Einstein equations give the anisotropic energy-momentum tensor, with the radial pressure the negative of the energy density, and the angular pressures being equal. Consequently, specifying the mass function simultaneously fixes the spacetime geometry and matter distribution. The resulting metric satisfies the curvature condition required for the massless Klein--Gordon equation separation and enabled derivation of bounds on the photon ring radius and the Lyapunov exponent using the induced temperature.

    The bounds on the photon ring radius were obtained for a constant equation-of-state parameter on the equatorial plane. The finiteness of the asymptotic mass requires $w > 1/2$, whereas the nonnegative energy density constrains the integration constant in the mass function to be nonnegative. The null, weak, and strong energy conditions were satisfied in the equatorial plane, whereas the DEC set an additional restriction $w \leq 1$. This mass function was nondecreasing and approached its asymptotic limit, and an outer horizon requires the rotation parameter to be at most the asymptotic mass. The photon ring radius of rotating black holes with anisotropic matter is bounded above by the Kerr photon ring radius for co-rotating and counter-rotating motions with identical asymptotic mass and rotation parameters. The lower bounds on the photon ring radius were obtained using the outer horizon radius and rotation parameter, with a further dependence on the energy density at the outer horizon for the co-rotating motion. Hence, the photon ring radius with anisotropic matter was constrained by the finite lower and upper bounds and does not exceed the Kerr photon ring radius.

    The Lyapunov exponent at the equatorial photon ring was derived from the second radial derivative of the effective potential for null geodesics and is measured relative to coordinate time. The upper bound on the Lyapunov exponent was obtained with anisotropic matter. The bound became an equality in vacuum to recover the Kerr results. The massless scalar field quasinormal mode frequency was evaluated within the first-order WKB approximation with the boundary conditions of purely ingoing waves at the outer horizon and purely outgoing waves at spatial infinity. In the eikonal limit, geometric optics relates the scalar modes to null geodesics. For the modes corresponding to equatorial photon rings, the absolute magnitude of the imaginary part of the quasinormal mode frequency is proportional to the Lyapunov exponent. Therefore, the upper bound on the Lyapunov exponent imposes an upper bound on the absolute magnitude of the imaginary part of the quasinormal mode frequency at each fixed overtone number in rotating spacetime with anisotropic matter.

    The induced temperature was analyzed for a probe string in plane-wave geometry derived along the equatorial photon ring. A parallel-propagated frame was built from the Killing--Yano tensor, and the Penrose limit along the photon ring yielded plane-wave geometry. The equations of motion for a probe string embedded into the plane-wave geometry followed from the Nambu--Goto action. The solution yielded an induced worldsheet metric with Rindler geometry. The components of $A_{ij}$ encode the tidal forces on neighboring null geodesics, and its trace is given by the negative null contraction of the Ricci tensor. The difference between the squared worldsheet surface gravity and the squared Lyapunov exponent is written in terms of the Ricci tensor null contraction and the contribution of matter to the spacetime curvature at the photon ring. The surface gravity of the Rindler worldsheet horizon determines the Unruh temperature, which is defined as the induced temperature, and provides an upper bound on the Lyapunov exponent $2 \pi T_\mathrm{ind} = \kappa_\mathrm{ind} \geq \lambda$. In the eikonal limit, the induced temperature provides bounds on the absolute magnitude of the imaginary part of the quasinormal frequency and the fundamental mode damping time, with corrections that depend on the matter distribution at the photon ring.

    Therefore, the finite asymptotic mass and nonnegative energy density bound the photon ring radius and Lyapunov exponent for rotating black holes in anisotropic matter. For a constant equatorial equation-of-state parameter, the finite asymptotic mass imposes $w > 1/2$ and the DEC restricts the range to $1/2 < w \leq 1$. The photon ring radius is at most the Kerr photon ring radius for identical asymptotic mass and rotation parameters, $r_{\PR}^{\pm} \leq r_{\PR,\mathrm{K}}^{\pm}$, whereas the lower bounds depend on the quantities at the outer horizon. This bound $2 \pi T_\mathrm{ind} = \kappa_\mathrm{ind} \geq \lambda$ extends the equality between the Lyapunov exponent and induced worldsheet surface gravity in vacuum Kerr spacetime to an inequality with anisotropic matter. This inequality has the same form as the Maldacena--Shenker--Stanford (MSS) bound for thermal quantum systems and the Hashimoto--Tanahashi (HT) bound for particle motion near black hole horizons. The MSS bound limits the OTOC growth exponent using the temperature of the quantum system, whereas the HT bound limits the Lyapunov exponent of the particle motion using the Hawking temperature. The induced temperature offers the associated thermal bound on the photon ring Lyapunov exponent. An extension of the analysis beyond the equatorial plane and the applicability of the bounds for a broader null geodesic motion remain for future work.

\section*{Acknowledgments}
    This research was supported by Basic Science Research Program through the National Research Foundation of Korea (NRF) funded by the Ministry of Education (NRF-2022R1I1A2063176) and the Dongguk University Research Fund of 2026.

\printbibliography

\end{document}